# Digital Twins and Blockchain for IoT Management

Mayra Samaniego[†]
Computer Science
University of Saskatchewan
Saskatoon Canada
mayra.samaniego@usask.ca

Ralph Deters
Computer Science
University of Saskatchewan
Saskatoon Canada
deters@cs.usask.ca

## ABSTRACT

Security and privacy are primary concerns in IoT management. Security breaches in IoT resources, such as smart sensors, can leak sensitive data and compromise the privacy of individuals. Effective IoT management requires a comprehensive approach to prioritize access security and data privacy protection. Digital twins create virtual representations of IoT resources. Blockchain adds decentralization, transparency, and reliability to IoT systems. This research integrates digital twins and blockchain to manage access to IoT data streaming. Digital twins are used to encapsulate data access and view configurations. Access is enabled on digital twins, not on IoT resources directly. Trust structures programmed as smart contracts are the ones that manage access to digital twins. Consequently, IoT resources are not exposed to third parties, and access security breaches can be prevented. Blockchain has been used to validate digital twins and store their configuration. The research presented in this paper enables multitenant access and customization of data streaming views and abstracts the complexity of data access management. This approach provides access and configuration security and data privacy protection.

## CCS

• Security and privacy → Systems security → Distributed systems;
• Information systems → Data management systems

[†] Corresponding Author

## 1 Introduction

Digital twins enable data transmission between physical and virtual environments [1]. They facilitate the monitoring, understanding, and optimization of physical resources. Digital twins became popular in the 2000s when used in the manufacturing industry. General Electric (GE) designed twins of their equipment to process data from sensors [2], [3]. Digital twins have also been used in different fields, such as industry [4], healthcare [5], [6], smart cities [7], [8], and agriculture [9], [10]. Digital twins create virtual representations of IoT resources [11]. A general overview of this representation is presented in Figure 1.

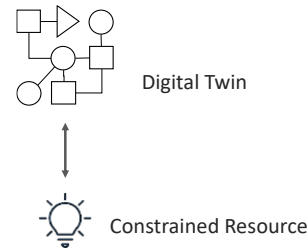

**Figure 1: General representation of a digital twin in IoT.**

Blockchain is a distributed ledger technology that enables direct transactions without central verification authorities [12]. Blockchain's design pattern and features can enable secure decentralized systems for IoT [13]. The main features of blockchain are a peer-to-peer network, cryptographically secure, append-only, immutable, and updatable via consensus.

An IoT architecture integrates at least three layers, perception, network, and application [14]–[16] (Figure 2). The perception layer represents smart sensors and actuators that collect data and interact with the environment. They are essential for operation and automation. The network layer handles communication among devices and nodes, controls data streaming from sensors to processing services, and ensures interaction among devices and the world. The application layer handles the provisioning of services to users. These services are interconnected and provide meaningful information, trigger actions, and support decision-making in real-time [15].

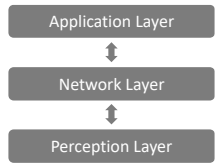

**Figure 2: Main layers of an IoT architecture.**

Regarding IoT management, security and privacy are open issues [17]. Privacy should be protected when accessing data by allowing individuals to control which of their data is collected, by whom, and when. Even though these open issues were presented in 2010, they are still considered security and privacy IoT challenges [18].

The research in this paper presents a system integrating digital twins and blockchain to manage access to data streaming in IoT. Digital twins are implemented to prevent third parties from directly accessing IoT resources. Access to data streaming is managed through digital twins. This approach provides resource security. Data streaming views are customized within digital twins. Every third party accessing data streaming has its digital twins customized individually. Therefore, anytime data streaming access for a specific third party needs to be updated, enabled, or disabled, the digital twins that belong to that third party are updated or turned on and off without affecting IoT resources (e.g., smart sensors) or other digital twins. This approach protects the privacy of individuals.

## 2 Background

### 2.1 Digital Twins and IoT

Digital twins are digital representations of physical or virtual entities [12]. The main characteristic of digital twins is that they are constantly updated and can implement their own communication methodology. For instance, Microsoft Azure has developed a digital twin definition language (DTDL) [19], combining APIs and JSON components. Additionally, Microsoft's digital twins' definition has a unique identification, integration of sensors and actuators to replicate the senses of physical resources, integration of some artificial intelligence (AI) capabilities to make some decisions on behalf of physical resources, trusting mechanisms to be trusted by the physical resources they represent and other digital twins, privacy and security to protect the state of the resources they represent, and communication capabilities to interact with physical resources, the physical environment, and other digital twins.

Certain concepts are sometimes linked to digital twins but are different, for instance, simulations, digital shadows, and digital threads.

- The main difference between simulations and digital twins is the update[20]. Digital twins are constantly updated, and simulations are not. Also, while digital twins enable a dynamic two-way data flow in cyber-physical spaces, simulations are static.

- The main difference between digital shadows and digital twins is the direction of data flow [21]. Digital shadows implement a one-way data flow between the virtual representation and the physical asset, mirroring the status of physical assets. Digital twins implement a bidirectional data flow between the virtual representation and the physical asset, enabling constant communication.

- The main difference between digital threads and digital twins is the state's recording [22]. Digital threads record the evolution of the physical asset over time. Digital twins do not record their state changes.

In an IoT ecosystem, access to data streaming from smart sensors at the perception layer is required to perform analysis, support efficient decision-making, and provide services. A smart city is an example of an IoT ecosystem. According to Ptak A. [23], a smart city aims to make citizens aware of their electricity service consumption patterns to make better consumption decisions. In this scenario, different stakeholders require access to data. Citizens require access to processed data through visual reports about their service consumption, and service providers need access to data streams to generate service consumption reports. Figure 3 shows an example of a smart city without digital twins. In this figure, stakeholders directly access sensor data streams. Figure 4 shows an example of a smart city that integrates digital twins to provision data streaming views. In this figure, stakeholders do not directly access sensor data streams. Instead, stakeholders access digital twins. Digital twins can be used in other IoT ecosystems, such as health care, supply chain, and smart agriculture. The smart city scenario is an example to provide a better understanding for readers.

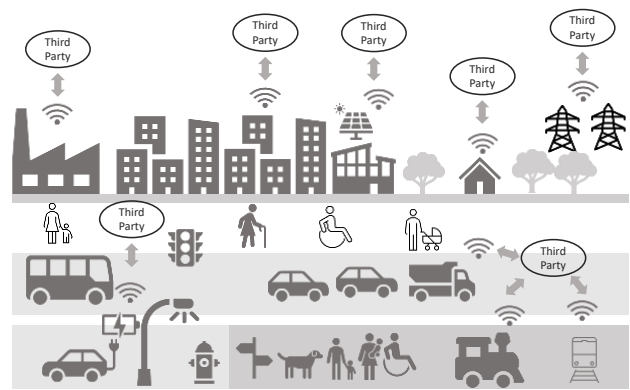

**Figure 3: Example of a smart city representation without digital twins**

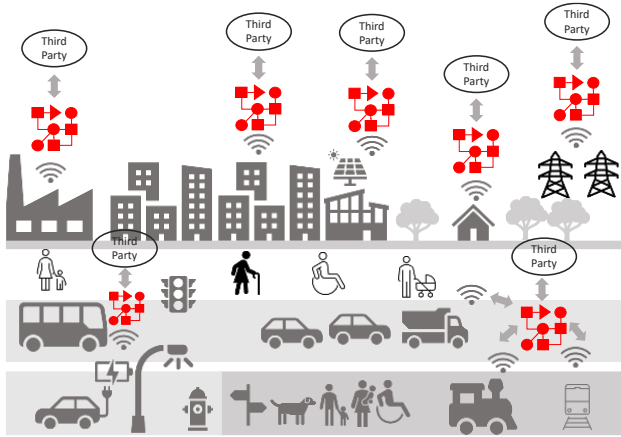

**Figure 4: Example of a smart city representation with digital twins.**

## 2.2 IoT Access Management and Privacy

Traditional data access management systems start from the premise that there is a data owner who owns a piece of data, and this piece of data can be shared with different users. In IoT, the premise is different. There are no data owners but device owners. There is no ownership over a piece of data but over the devices that stream data. IoT is a dynamic environment in which data streams are generated and communicated differently than traditional internet applications. Table 1 summarizes data interactions in IoT and internet systems.

**Table 1: Example of IoT interactions vs. traditional internet interactions**

| Parameter | Traditional Internet | IoT |
| --- | --- | --- |
| Data creation | By humans | By sensors |
| Data consumption | By request (search engines) | By pushing information and triggering actions |
| Data flow | Through links that connect web resources | Through defined operators that connect devices and constrained services |
| Data Value | Answer users' questions | Action and timely information. The value is added by analytics. |

Role-based access control model (RBAC) [24], attribute-based access control model (ABAC) [25], and capability-based access control model (CAC) [26] have been used for access management in IoT.

- RBAC defines roles to manage access and grants rights to roles, not to users [24].
- ABAC defines attributes that users have to prove to get access rights. ABAC provides higher flexibility because users do not have to define their roles a priori explicitly. However, all participants must agree on the attributes used to validate access [27].
- CAC states that a specific token authority drives access rights. Tokens are unforgeable, communicable, and transferable [28]. The token owner will always be allowed to execute all the actions on the resource [28]. A token is linked to a set of access rights, represented as a capability [29]. The capability-based security approach would be a better fit for the distributed nature of IoT because it provides fast and effortless usability and high flexibility [30].

The following studies implement RBAC, ABAC, and CAC access control models in IoT. Kim et al. [31] propose an RBAC model to preserve sensitive data privacy in pervasive environments, no matter the changes that might occur in access control permissions. The authors define nine components to establish the level of access to a role: subject, privacy role, privacy permission, session, constraints, user assignment, role hierarchy, context rule assignment, and context permission assignment. Chen et al. [32] present an RBAC access control system that, instead of centralizing roles management, integrates IoT servers that cooperatively support the access control components, reduce internal security threats, and execute a local evaluation of trust based on the roles of users and constrained devices. Bezawada et al. [33] propose the utilization of ABAC to guarantee security in IoT smart homes. The authors provide a list of attributes that fit the requirements of IoT home environments. Bhatt et al. [34] propose an ABAC model to control access and communication in IoT. Instead of defining attributes of entities, the authors define attributes of the communication between entities to secure the communication flow between nodes. Neto et al. [35] present an ABAC model to control access to IoT home devices to manage the access control scheme to handle constrained device operations. Gusmeroli et al. [36] present a CAC system to secure access to IoT house appliances. The authors present a six-step process: checking the token's validity, checking the granted action, checking the conditions to execute the token, checking the validity of the signature, and checking the legitimation of the user.

IoT privacy has been the subject of some research. A survey by Sarwar et al. [37] presents taxonomies, issues, and trends regarding IoT privacy. A survey by Ogonji et al. [38] discusses privacy and security concerns in IoT and a threat taxonomy. Kong et al. [39] present a data-sharing scheme for the internet of vehicles that preserves privacy. Xiong et al. [40] present a framework for connected autonomous vehicles to exchange sensor data preserving privacy. Chanson et al. [41] propose a blockchain-based data protection and certification system. In IoT, access has been generally enabled directly over IoT resources, compromising security. Security breaches in constrained resources can leak sensitive data and risk the privacy of individuals. Table 2 summarizes publications that address access management at the IoT perception, network, and application layers.

Table 2: Publications that address access management at IoT's perception, network, and application layers.

| | |
|---|---|
| Atzori et al. [17] | **Perception**: addressing and naming, authentication, configuration.<br>**Network**: mobility, data integrity, protocols, traffic characterization<br>**Application**: data privacy, digital forgetting, interoperability of services, scalability, decentralization |
| Mahalle et al. [42] | **Perception**: addressing and naming<br>**Network**: N/A<br>**Application**: secure interaction among devices, users, and services |
| Shancang et al. [43] | **Perception**: interoperability<br>**Network**: protocols<br>**Application**: service discovery, integration with ICT systems, data trustworthiness, data protection, data privacy |
| Gubbi et al. [44] | **Perception**: energy efficiency, availability<br>**Network**: N/A<br>**Application**: data privacy, GIS-based data visualization, data mining |
| Borgia E. [45] | **Perception**: addressing and naming, mobility, interoperability, configuration<br>**Network**: M2M communication, end-to-end reliability, secure bootstrapping of objects, data transmission, traffic characterization<br>**Application**: big data, data processing and analysis in real-time, data scalability, ethical and privacy issues, secure access to data by authorized services, security of IoT data |
| Al-Fuqaha et al. [46] | **Perception**: fault, configuration, accounting, performance and security (FCAPS), availability, scalability, interoperability<br>**Network**: mobility<br>**Application**: reliability of services, availability of services, scalability, interoperability, data security, data privacy |
| Farhan et al. [47] | **Perception**: scalability, interoperability, energy efficiency, fault tolerance<br>**Network**: N/A<br>**Application**: data privacy, self-capabilities, software development |
| Khan et al. [48] | **Perception**: energy efficiency, authentication, authorization, accounting<br>**Network**: N/A<br>**Application**: data privacy, data confidentiality, data integrity, authentication, authorization, accounting, availability of services, single points of failure |

## 2.3 Blockchain and IoT

Blockchain has been studied as a means to design management systems for IoT, such as [49]–[51]. Gbadebo et al. [52] propose a blockchain-based system to manage data about IoT devices and access to devices. Fan et al. [53] propose a prototype that uses a permissioned blockchain to store device data and validate device transactions. Thakker et al. [54] present a blockchain-based system to give authenticity certificates for IoT devices. Hasegawa et al. [55] and Vinay et al. [56] present blockchain-based systems to manage the authenticity of data at the perception layer. Cui et al. [57] propose a blockchain-based solution to manage the interoperability of trusted devices and restrict devices tagged as untrusted. Jiang et al. [58] propose integrating multiple blockchain technologies to manage device identification and data storage. Zhaofeng et al. [59] present a blockchain-based data management scheme solution for IoT data management at the edge level. Truong et al. [60] present a blockchain-based system to store access control policies that allow data owners to receive remuneration for their data. Laurent et al. [61] present a blockchain-based solution that allows data owners to define access rights for their data. Lu et al. [62] propose a blockchain collaborative architecture to enable data sharing for multiple parties. Manzoor et al. [63] present a blockchain system to sell sensor measurements to different users. Si et al. [64] propose a blockchain-based framework to handle security when sharing IoT data. Sultana et al. [65] present a data-sharing system that integrates smart contracts to manage access control and enable communication among IoT devices. Liu et al. [66] propose a data-sharing system for mobile terminals. Rahman et al. [67] propose a blockchain-based infrastructure to support the sharing economy in smart cities. Agyekum et al. [68] present the design of a data-sharing module for IoT that uses blockchain as a proxy server to re-encrypt data. Blockchain has also been studied for IoT management in healthcare [69], [70], industry [71][72], autonomous management [73], zero-trust management [74], configuration management [75], [76], multitenant access control management [77], and detection of suspicious operations [78].

## 3 Digital Twins and Blockchain to Manage Access to IoT Data Streams

Blockchain technology can help researchers refine the definitions of digital twins by creating a transparent industry environment and proof of their legitimacy and identity [79]. Pethuru R. [80] highlights the cryptography benefits of defining digital twins with blockchain. Suhail et al. [81] review how blockchain has been used to design digital twins and present a framework that integrates blockchain for data synchronization and storage.

The research presented in this paper implements digital twins to abstract and customize access to IoT data streaming. Therefore, third parties do not access IoT resources but digital twins. Digital twins encapsulate individual access and data streaming view configurations, which provide third parties with specific permissions and data views when accessing the same IoT resource. As a result, multitenant access is enabled. In this research, digital twins have been implemented as REST (representational state transfer) [82] microservices and their configuration is stored in-chain.

Digital twins' access management presented in this research follows the trust institution presented by Lilian Edwards [83]. Edwards describes a trust as an institution with three participants, the settlor, the trustee, and the beneficiary.

- The settlor owns property over an asset (any asset) and transfers that property to the trustee.

- The trustee exercises rights after that as the owner of these assets but is not entitled to use the trust assets as their own. The terms of the trust restrict the use of the trust asset.
- The beneficiary receives the benefits of the property.

Edwards presents the idea of considering the settlor as the trust's main beneficiary. Therefore, the trustee owes fiduciary duties to the settlor. These roles create trust relationships over any asset. Edwards presents an example of creating a trust relationship over web data (Figure 5). Through Edwards' trust institution, this research creates trust structures representing the property transfer between settlors and trustees:

- Each device owner becomes a settlor.
- Each third party willing to access IoT data streaming becomes a trustee.
- The settlor transfers ownership of digital twins to the trustee.
- Each trustee gets access only to their digital twins.

The access management solution presented in this research follows the trust institution presented by Lilian Edwards, but instead of using it to manage access to web data, the trust institution is used to manage access to digital twins (Figure 6). The system architecture presented in this research integrates digital twins, trust structures, and blockchain technology to fulfill Edwards' trust institution.

Figure 7 shows the overall architecture of the system. The system seeks to enable decentralized, transparent, reliable, and customizable data streaming views for different trustees individually. This research integrates blockchain to provide decentralization, transparency, and reliability to the system and the provisioning of digital twins. Blockchain is the entity placed between the settlor and the trustee to perform the following tasks:

- Twin storage
- Twin validation
- Twin provisioning to the proper IoT resources
- Twin provisioning to the proper trustees
- Access validation to digital twins
- Trust structure validation
- Trust structure storage
- Property transfer validation
- Settlor, trustee, and twin matching verification

The definition of digital twins presented in this research encapsulates the following configurations for data streaming views:

- Settlor hash identification
- Trustee hash identification
- Twin hash identification
- Data streaming start
- Data streaming ending
- Data view parameters

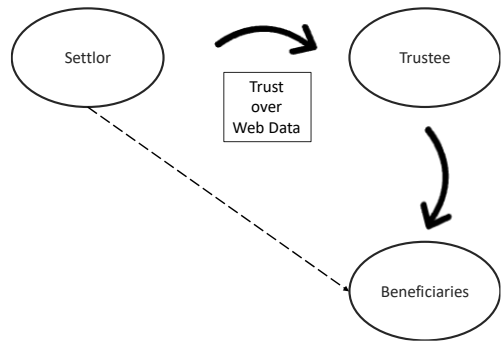

**Figure 5: Trust institution over web data.**

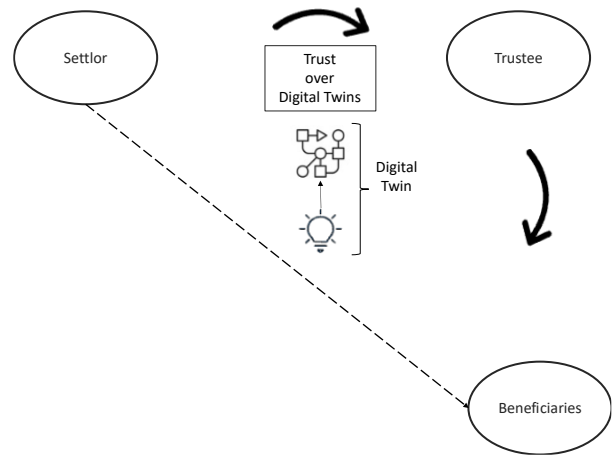

**Figure 6: Trust institution over digital twins.**

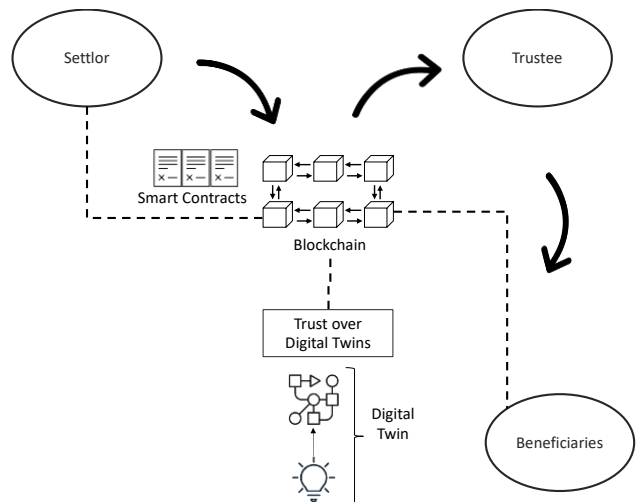

**Figure 7: Overall architecture of the presented system with digital twins and blockchain.**

Also, the definition of digital twins presented in this research encapsulates the following functionalities:

- Settlor identity verification
- Trustee identity verification
- Twin identity verification
- Twin communication
- Resource communication
- Data view provisioning

In the initial example of a smart city, the actors of the trust structure are defined as follows:

- The settlor is the smart city government
- The asset is the digital twins
- The trustees are the parties that provide data visualization services to citizens

There can be different specifications for the previously listed actors depending on the level of pervasiveness. The fact of integrating blockchain in the implementation of digital twins and the establishment of trust structures to access them guarantees security in the following aspects:

- IoT resources get an access security layer because all intended access is first validated by digital twins. Also, data streams are provisioned by digital twins.
- IoT resource owners get privacy protection because they can detail who gets access to their data streams and how that data is viewed.
- IoT access management gets a mechanism to enforce access compliance and data stream views customized separately for different third parties.

### 3.1 Implementation and Experiments

This research develops digital twins using the Golang programming language [84]. Figure 8 shows an example of a Golang function that starts the operation of a digital twin exposing the communication interface. The digital twins developed in this research expose the following APIs:

- Constrained application protocol (CoAP) [85] API to handle communication among digital twins. Multiple digital twins communicate with each other to provide customized data streaming views among them if needed.
- Hypertext transfer protocol (HTTP) [86] API to handle communication between digital twins and blockchain to retrieve twins' configurations.
- HTTP API to handle communication with third parties wanting to retrieve data streaming.
- Figure 9 shows examples of these exposed APIs.

```go
func main() {
    go start_local_block_building()
    /* listen all uncome temperatures */
    coap.HandleFunc("/listen", func(r
    *coap.RemoteAddr, m *coap.Message) *coap.Message
    {
        handle_request (m)
```

**Figure 8: Golang routine that starts a digital twin.**

```
coap://twin001/coap_api/talk_to_dt
http://twin001/http_api/talk_to_bc
http://twin001/http_api/talk_to_third_party
```

**Figure 9: Examples of CoAP and HTTP APIs exposed by digital twins.**

In this research, an Ethereum private blockchain network in a fog layer is deployed. The fog layer is formed by three servers with 4GB RAM, 250 GB storage, and Ubuntu 20.05 operating system. Trust structures are implemented as smart contracts programmed in the Solidity programming language [87]. The digital twins' configurations are stored in-chain as smart contract structures for verification and transparency. Figure 10 shows some functions for the defined digital twins' in-chain configuration. This research follows two programming styles, logs and variables, to evaluate the cost impact on the digital twins hosted in a fog layer. Figure 11 shows an example of a trust function programmed using logs. Figure 12 shows an example of the same trust function programmed using variables.

- Logs store data as part of transaction receipts. They are generated by the programmed clients when executing transactions and stored alongside the blockchain to allow retrieving them. Variables are programmed within smart contracts.
- Logs can help save costs as their operations represent 375 units of gas, while a regular Ethereum transaction starts at 21000 units of gas [88].

Web3.js is a group of libraries that allow interaction with an Ethereum network [89]. The components enabling in-chain and off-chain communication are programmed in NodeJS programming language [90] and web3.js. These components enable reading and writing in-chain transactions. We used virtual resources to create a test environment at the perception level [75], [91]–[93]. Therefore, IoT-constrained devices are simulated.

```
enum ViewFormatChoices {JSON, XML}
ViewFormatChoices constant default_format_choice=
ViewFormatChoices.JSON;

struct DigitalTwin {
    string twin_id;
    address twin_setlor;
    address twin_trustee;
    uint256 streaming_start;
    uint256 streaming_end;
    DataView streaming_view;
}

struct DataView{
    uint256 streaming_period;
    ViewFormatChoices view_format;
}

DigitalTwin[] digital_twins_farm;

function getDigitalTwin(uint index) public view returns
(DigitalTwin memory) {
    require(digital_twins_farm[index].twin_setlor==msg.sender);
    return digital_twins_farm[index];
}

function setDigitalTwin(string memory _twin_id, address
_twin_setlor, address _twin_trustee, uint256
_streaming_start, uint256 streaming_end) public {
    DataView memory dv;
    digital_twins_farm.push(DigitalTwin(_twin_id,
    _twin_setlor,_twin_trustee,
    _streaming_start,streaming_end, dv));
`
```

**Figure 10: Smart contract interface for the configuration of digital twins.**

```
event EventTwinAccessRegistration(
    string twin,
    address indexed t_settlor,
    address indexed t_trustee
);
```

**Figure 11: Trust function programmed with logs.**

```
struct TwinAccessStruct{
    string twin;
    address t_settlor;
    address t_trustee;
}

TwinAccessStruct dos;
```

**Figure 12: Trust function programmed with variables.**

The results obtained from evaluations help us understand the impact of blockchain-based REST digital twins as an access security layer for IoT and privacy protection for individuals. Also, these experiments help us understand the cost of provisioning customized digital twins. This research assesses the cost in terms of gas consumption of the two previously detailed programming styles, variables and logs. Evaluating costs is essential due to the number of constrained devices an IoT ecosystem can have, like the example of a smart city. Therefore, anything that helps save costs is welcome. Figure 13 shows the result of the deployment cost of the smart contract interface that handles digital twin configurations using variables and logs. Deploying digital twins using variables costs 418412 units of gas while deploying them using logs costs 229533 units. Deploying digital twins as transactional logs from contract events represents 45.14% less gas consumption.

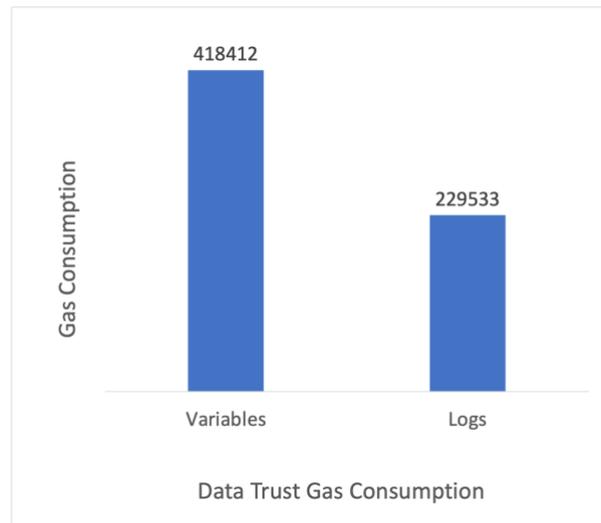

**Figure 13: Deployment cost of digital twins.**

Figure 14 shows the results of the in-chain storage cost of digital twins using variables and logs. The parameters stored were the settlor's hash account, the trustee's hash account, and the digital twin's hash value. Storing digital twins as variables costs 42773 units of gas while storing them as transactional logs costs 29012 units of gas.

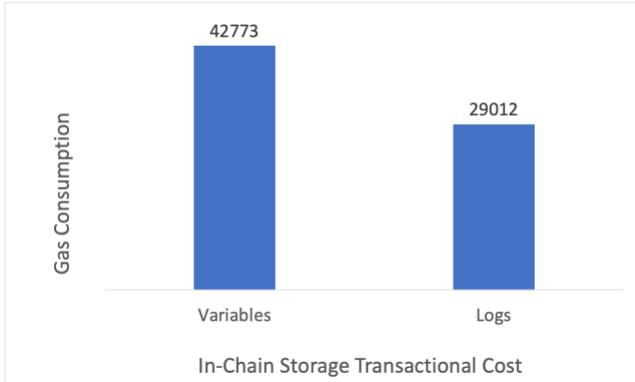

**Figure 14: Storage cost of digital twins.**

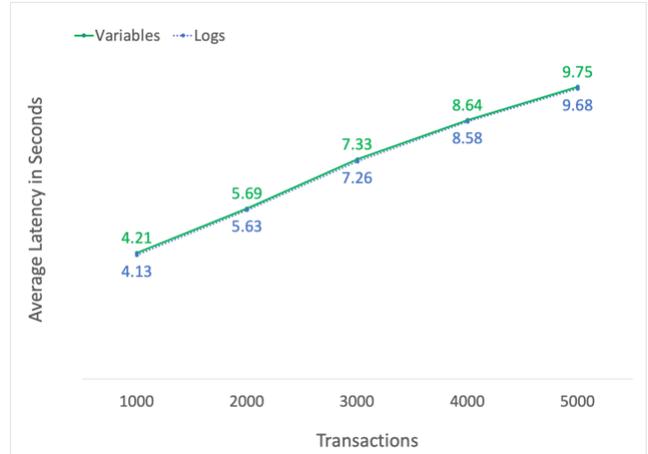

**Figure 16: Average system performance per transaction when creating digital twins.**

The following experiment evaluates the performance of the system when creating different loads of digital twins. Different groups of requests were sent, from one thousand to five thousand. Figure 15 shows the latency when creating the whole group of digital twins (1000 to 5000), and Figure 16 shows the average latency per transaction. Creating digital twins with logs produces a slightly lower latency than variables. On average, it is 1% less latency. As expected, the more requests are sent, the more latency the system generates. The results shown in Figures 15 and 16 also show that no matter the programming style, the latency is high. This is directly related to the blockchain consensus mechanism that is implemented in this research. In this research, a private Ethereum network with a proof-of-work (PoW) consensus [94] is implemented. PoW requires that miners compete to solve a hash using their computer processing power. Pow is considered the most secure consensus mechanism as its features combat attempts to duplicate spending. However, that security obtained sacrifices performance.

## 4 Conclusions

This research presents a system integrating digital twins and blockchain technology to manage access to IoT data streaming. Digital twins are implemented to encapsulate the configuration of individual data streaming views for different parties. Therefore, each third party gets digital twins with a specific access configuration to view the data from the same IoT resource. This approach enables multitenant access. Our system manages access to digital twins instead of access to IoT resources. Access to digital twins is managed through trust structures in the form of smart contracts that follow the trust institution presented by Lilian Edwards.

Changes or updates to data streaming views are done over the digital twins without directly accessing IoT resources. IoT resources streaming data are protected from unauthorized access since digital twins validate identities and permissions and provision data streaming. This approach adds security in terms of access and configuration. Additionally, through digital twins, users can control who can access their data streaming and how it is provisioned. This approach protects individuals' privacy.

Blockchain technology is integrated into our system because its features add transparency and reliability to validate and store the configuration of digital twins and allow the creation of trust structures with smart contracts to access them. Smart contracts make digital twins and trust structures immutable, self-verifying, self-enforcing, self-executing, and tamper-proof. Therefore they can be used as auditing mechanisms. This research presents a smart city use case example. In this example, neither the government nor the service providers are involved in validating digital twins and trust structures, but rather the blockchain network is. Trust structures can be designed following different architectural styles, not only a blockchain-based approach. For instance, a centralized client-server architecture. However, this option does not provide transparency and reliability because a single entity would handle validation without additional verification.

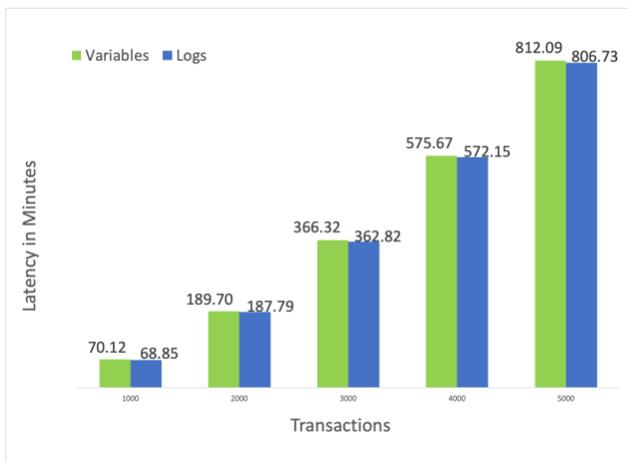

**Figure 15: System performance when creating digital twins.**

This research evaluates the gas cost of deploying and storing digital twins. Digital twins are programmed with two solidity programming styles, logs and variables. These programming styles refer to how data is stored in a smart contract. Logs have not been explored as much as variables. The default programming style is variables. Logs offer gas saving, but they are not accessible from other smart contracts but only by events and can store up to three indexed parameters (indexed parameters are searchable). This does not affect the system presented in this research since there is no interaction between smart contracts, and the data from the trust structures are accessed through events. If storing more than three indexed parameters is desired or needed, multiple parameters can be hashed into one hashing value and stored in-chain, or they can be stored as non-indexed log data. The latter option affects the system presented in this research because searches cannot be implemented since non-indexed parameters are not searchable.

The experiment results show that the deployment saving is approximately 45% with logs. On the other hand, the system's performance is similar using either of the two methods because the difference is only 1% of latency. Based on these results, the selection of the programming style depends on the number of indexed parameters to store in-chain and the units of gas willing to be spent. Finally, experiment results indicate that the solidity programming style impacts the deployment cost but does not significantly impact the system's performance. What impacts performance is the blockchain consensus mechanism. The more security the consensus algorithm provides, the more latency the system has. In this research, security through PoW was chosen over performance. PoW is considered the most secure consensus mechanism. Other consensus algorithms can improve latency but would sacrifice some security. In this research, security was preferred over performance. This preference might change depending on the characteristics and needs of specific scenarios. If high performance is required, then another consensus algorithm must be selected, such as proof of stake (PoS) [95], proof of authority (PoA) [96], or practical byzantine fault tolerance (PBFT) [97]. Future work will implement experiments on a large scale. This will allow the simulation of more complex IoT ecosystems and the testing of the presented system's ability to handle higher amounts of connected resources, data, and workloads.